\documentclass[amssymb,amsfonts,aps,prb,twocolumn]{revtex4}
\usepackage{graphicx}

\begin{document}

\title{Electron Confinement Induced by Diluted Hydrogen-like Ad-atoms in Graphene Ribbons}

\date{\today} 

\author{J. W. Gonz\'alez,$^{\dag,\,\ddag,\,}$\footnote{Corresponding author: sgkgosaj@ehu.eus} L. Rosales$^{\ddag}$, M. Pacheco$^{\ddag}$, A. Ayuela$^{\dag}$}
\affiliation{
\dag Centro de F\'{i}sica de Materiales (CSIC-UPV/EHU)-Material Physics Center (MPC), Donostia International Physics Center (DIPC), Departamento de F\'{i}sica de Materiales, Fac. Qu\'{i}micas UPV/EHU. Paseo Manuel de Lardizabal 5, 20018, San Sebasti\'an-Spain
\\
\ddag Departamento de F\'{i}sica, Universidad  T\'{e}cnica Federico Santa Mar\'{\i}a, Casilla 110 V, Valpara\'{i}so-Chile.}


\begin{abstract}
We report the electronic properties of two-dimensional systems made of graphene nanoribbons 
which are patterned with ad-atoms in two separated regions. Due to the extra electronic confinement 
induced by the presence of the impurities, we find resonant levels, quasi-bound and impurity-induced 
localized states, which determine the transport properties of the system. 
Regardless of the ad-atom distribution in the system, we apply band-folding procedures to simple 
models and predict the energies and the spatial distribution of those impurity-induced states.
We take into account two different scenarios: gapped graphene and the presence of randomly distributed 
ad-atoms in a low dilution regime. In both cases the defect-induced resonances are still detected. 
Our findings would encourage experimentalist to synthesize these systems and characterize their 
quasi-localized states employing, for instance, scanning tunneling spectroscopy (STS). Additionally, 
the resonant transport features could be used in electronic applications and molecular sensor devices.
\end{abstract}

\maketitle

\section{\label{sec:intro} Introduction}

Two-dimensional nanostructures are promising candidates for novel application at the nanoscale. 
Due to its dimensionality, it is possible to manipulate the entire system through several 
experimental techniques. These nanosystems are suggested for sensing applications because they 
have large surface-volume ratio, high electronic mobility, and externally tuneable conductivity. 
One of its applications is related to their use in high-precision molecular and magnetic sensors, 
which requires robust states at well-defined 
energies.\cite{Lehtinen2003,schedin2007detection,candini2011graphene,pisana2009tunable,JW_NL13,JW_adsorbates} 
In 2D electronics, the control of the electronic and magnetic properties can be achieved by specific 
point defects, such as vacancies, Stone-Wales defects and doping with ad-atoms, among others. \cite{banhart2010structural,nanda2012electronic,chen2011tunable,Lin2014observation,Buchs2009} 
Previous theoretical works propose electron confinement by doping 2D graphene with hydrogen atoms, 
which is used to manufacture narrowed graphene ribbons, quantum dots or junctions without the need 
for cutting or etching the system.\cite{yang2010two,xiang2009narrow,chernozatonskiui2007superlattices} 
Furthermore, it was experimentally shown that the adsorption of hydrogen atoms on graphene can be 
controlled by scanning tunneling microscope to decorate with patterns the system, which remains stable 
even at room temperature.\cite{sessi2009patterning} 
%

\begin{figure}[hb!]
\includegraphics[clip,width=0.4\textwidth,angle=0,clip]{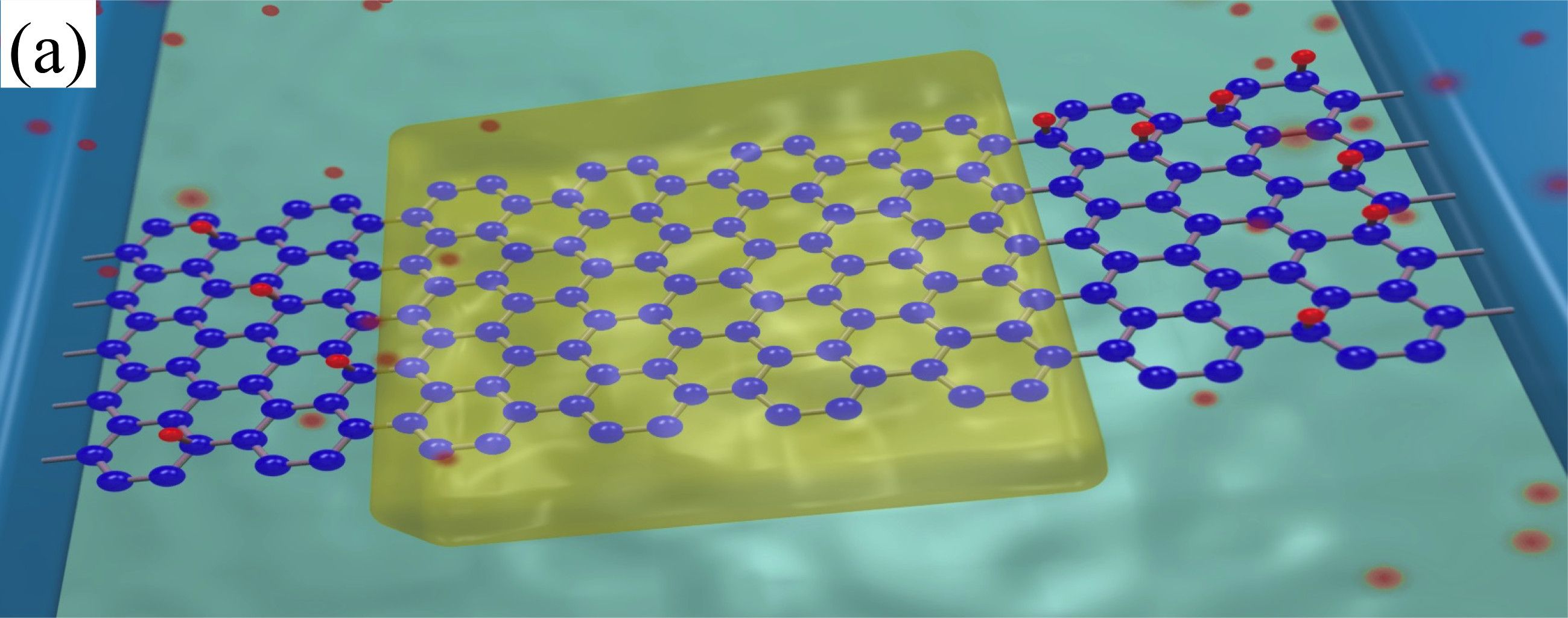} 
\includegraphics[clip,width=0.45\textwidth,angle=0,clip]{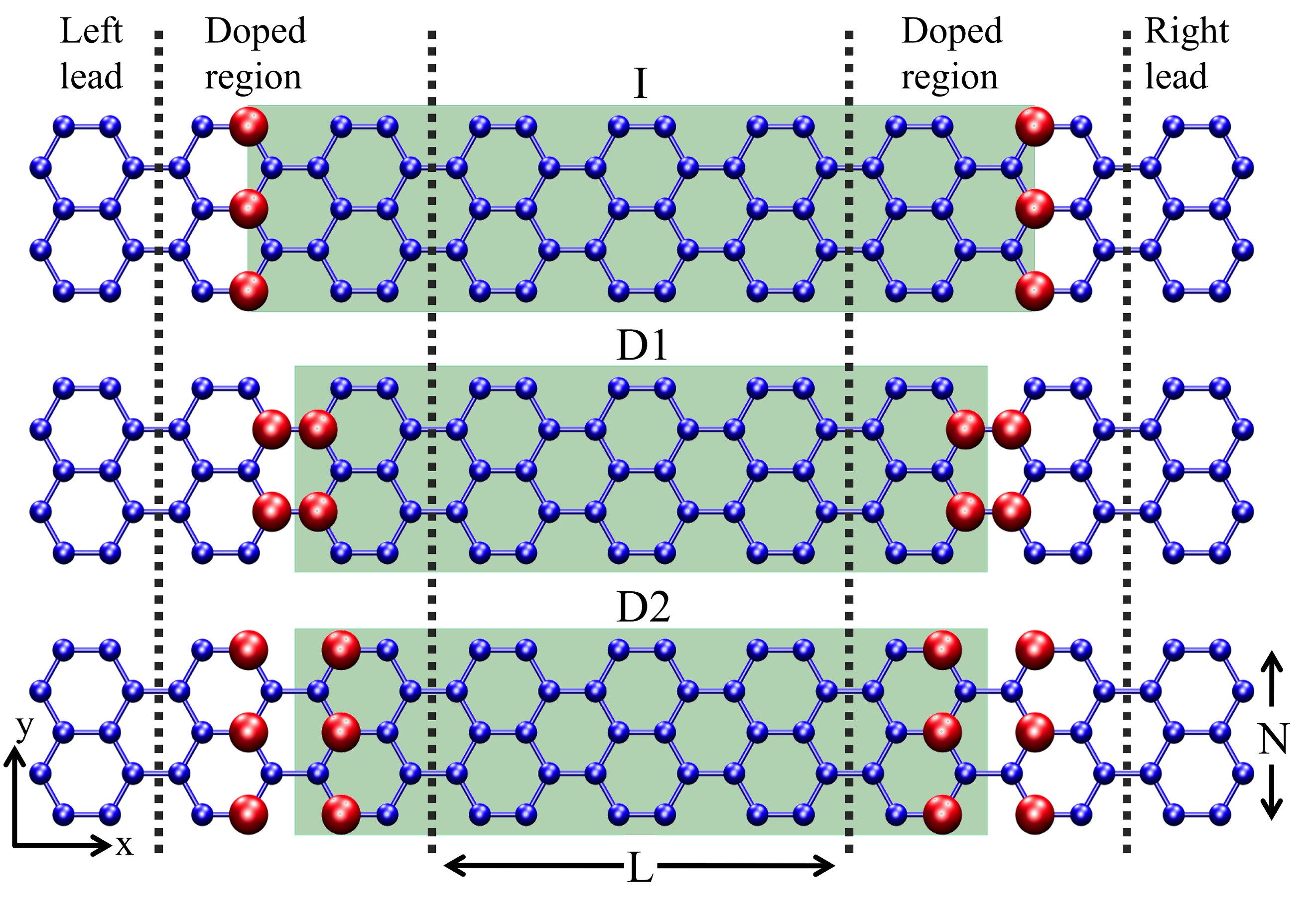} 
\caption{(Color online) (a) Proposed system based on GNR with ad-atoms in two separated regions. 
(b) Schematic view of ad-atoms in ordered configurations. The length $L$ (in units of $3$ $a_{CC}$) 
is the separation between the two doped regions.
The I and D notations stand for individual and double hydrogen column.
Red spheres represent hydrogen ad-atoms.
The effective system length $L_{eff}$ is given by the shaded region.}
\label{fig1:esq}
\end{figure}

Graphene nanoribbons ideally could be today synthesized by unzipping carbon nanotubes. This method allows provides narrow graphene nanoribbons suspended between Au (111) contacts or  deposited on a substrate such as SiC or BN. \cite{kosynkin2009longitudinal,tao2011spatially,jiao2009narrow,jiao2010facile} 
Such nanoribbons would be later protected from hydrogen deposition by a controlled coating, for instance, of PmPV (m-phenylenevinylene-co-2,5-dioctoxy-p-phenylenevinylene) at two different regions.\cite{jiao2010facile} 
The partial hydrogenation could be achieved by either direct exposure or exposing the sample to a beam of atomic H.\cite{balog2009atomic,scheffler2012probing}
Then, scanning tunneling microscopy (STM) and scanning tunneling spectroscopy (STS) could be used to characterize the 
quasi-localized states induced by the two H regions.\cite{kislitsyn2014vibrational}  
Note that STM experiments are able to control the H deposition in graphene sublattices \cite{Brihuegaprivate}. In this way, systems as those shown in Fig. \ref{fig1:esq} can be produced.

In graphene, the hydrogen-like ad-atoms are adsorbed on top of the carbon atoms through covalent 
bonds. The local hybridization of graphene changes from  $sp^2$ to $sp^3$-like, so that the 
low-energy conducting $\pi$-bands are affected.\cite{Boukhvalov_PRB08,ziatdinov2014direct} 
When hydrogen-like atoms are binding on top of carbon atoms, the graphene sublattice symmetry is 
broken, and thus, localized states appear near the Dirac point. These states at zero energy are well-known in carbon nanostructures\cite{Santos2012,ziatdinov2014direct,gonzalez2012impurity} 
and are comparable, in the low-energy range to a vacancy in $\pi$ models.\cite{Pereira_PRL96}

Charge carriers in graphene have a linear energy dispersion, then electron scattering obey 
the Klein's paradox.\cite{klein1929reflexion}  Klein showed that when the potential is of 
the order of the electron kinetic energy, the barrier for massless fermions becomes transparent. 
In practical terms, the Klein's paradox implies that carriers in graphene can not be confined 
by potential barriers.\cite{katsnelson2006chiral,young2009quantum} 
In this context, the question that motivates this work is the following: Is it possible to 
induce and control resonant states, extended across the whole system, by using impurities in graphene?.

In this paper we study the electronic and transport properties of graphene ribbons patterned with hydrogen-like ad-atoms at two separated regions, as it is shown in Fig.\ref{fig1:esq}. This approach can be used  as a model for any singly covalent adsorbed molecules in graphene. 
We are describing the H interaction on top of C atoms using a tight-binding Hamiltonian, which are  compared with density functional theory (DFT) calculations. 
We calculate the local density-of-states (LDOS) and the two-terminal conductance of the system, and we are considering the effects of the H-adsorbate impurity density, the spatial distribution and the sublattice balance at the doped regions. We find not only the expected impurity-induced localized states, but other kinds of electronic states such as resonant and interface states. These levels, which determine the transport properties of the system, appear at well-defined energies that we can unravel, and additionally infer their spatial distribution, by applying band-folding models. 

To bring our results into contact with experiments, we examine a random distribution of hydrogen adsorbate atoms at low concentrations in the two doped regions. The average LDOS still exhibits a series of defined peaks at some energy values, which behave similarly to the ordered ad-atom arrangements. We  are also study a ribbon over a substrate, modeled by a constant staggered potential (a mass term in the Dirac equation) which open a gap in the LDOS and conductance curves.  In the latter system, the behavior of the electronic states is still be determined by applying the same kind of simple folding models. 

Note that our proposal goes further than the existing reports, which have considered  periodic boundary conditions and/or superlattice heterostructures, where the electronic properties and the  transport are calculated parallel to the confinement generated by barriers.\cite{chernozatonskii2007two,chernozatonskiui2007superlattices,yang2010two,xiang2010thermodynamically}

\section{\label{sec:Model} Theoretical Description of the System}
The low-energy  electronic properties of graphene are well described within a single $\pi$-band  approximation in a nearest-neighbor tight-binding scheme. The Hamiltonian of the system can be written as
\begin{equation}\label{H}
\mathcal{H} =  - {\gamma _0}\sum\limits_{\left\langle {l,m} \right\rangle } {c_l^\dag {c_m}}  + 
{\varepsilon _H}d_n^\dag {d_n} - {\gamma _H}\left( {c_{{p_n}}^\dag {d_n} + c_{{p_n}}^{}d_n^\dag } \right),
\end{equation}
where the hopping energy between nearest-neighbor carbon atoms is $\gamma _0=2.75$ eV, 
$c_m ( c_m^\dag )$ is the C annihilation (creation) operator on the $m$-th site of graphene lattice, 
$\gamma _H$ is the coupling energy between the carbon and the hydrogen ad-atom, $\varepsilon_H$ is the 
hydrogen on-site energy, $d_n \left( d_n^\dag\right)$ is the H annihilation (creation) 
operator on the adsorbate site, and $c_{{p_n}}$ is the host carbon atom bonded to the hydrogen atom.
We use the set of tight-binding parameters $\gamma _H = 5.72$ eV $= 2.08$ $\gamma _0$  
and $\varepsilon_H = 0$ determined from first-principles band structure calculations of hydrogenated graphene.\cite{Bang_PRB81,Schmidt_PRB81,Wehling_PRL2010}. It is worth to mention that the effects of considering $\varepsilon_H \neq 0$ are not being considered because preliminary tests show that this value split  energy states outside of the energy window studied in this paper.
Details about the calculations of the conductance and LDOS are presented in the Appendix \ref{Apex_cond}.
We focus on armchair nanoribbons, which at the edges mix sublattices and valleys, so that the spin the degree of freedom entails to have doubly degenerate bands, in contradistinction to zigzag ribbons.\cite{brey2006electronic}

\section{Results}

In this section we present results of LDOS and conductance of armchair graphene nanoribbons (AGNR) in the presence of ordered and random distributions of hydrogen ad-atoms in two separated regions of the ribbon. Besides, we explain the LDOS behavior by applying  folding of certain bands of simple effective models. The LDOS is projected in all atoms of the system, marked by the outer dashed lines in Fig. \ref{fig1:esq}.

\subsection{General Trends on the Local Density of States and Conductance}
We chose a metallic AGNR in the range of the ultra-narrow ribbons, recently synthesized. \cite{denk2014exciton} The conductance curves of these narrow ribbons exhibit broad plateaus at low energy,  making easier the visualization of the doping effects. Nevertheless, we would like to note that our findings in the  LDOS and conductance curves are general and appear in wider metallic and semiconductor ribbons. 
The considered separation between the two doped regions  is $L=15 = 45 \, a_{CC}$  and the AGNR width 
is $N=11 = 5\sqrt{3} \, a_{CC}$. In this nanoribbon, we have considered the three ordered arrangements of adsorbed hydrogen atoms following the scheme of Fig. \ref{fig1:esq}.

\begin{figure}[bh!]
\includegraphics[width=0.49\textwidth,angle=0,clip]{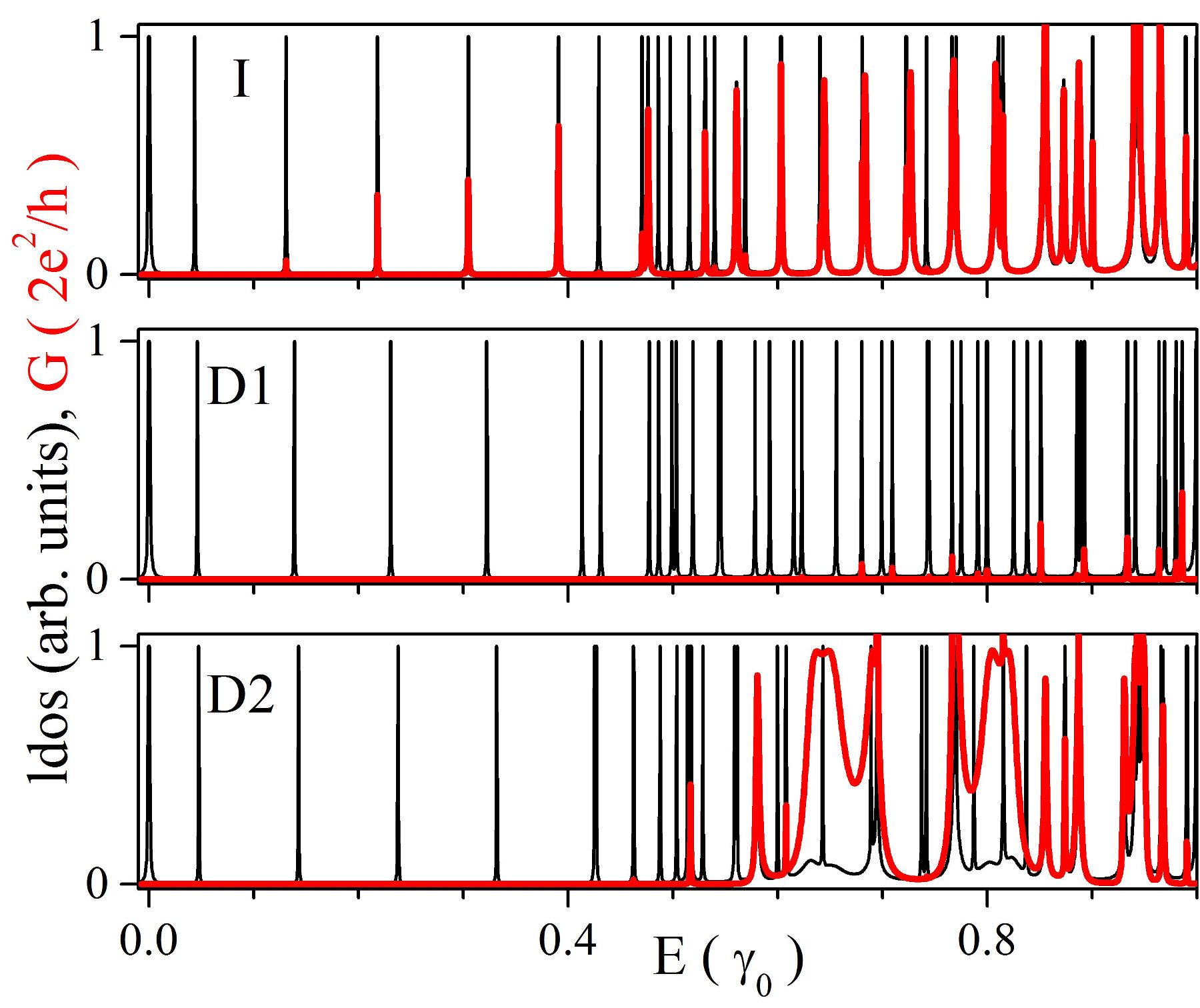} 
\caption{ (Color online) Local density of states (black line) and conductance (red line) as a function of energy for different hydrogen distributions, separated by a length $L = 15$  on a metallic AGNR $N = 11$.}
\label{fig2:GL_E}
\end{figure}

We start the analysis by examining the LDOS as a function of the Fermi energy, as it is shown in Fig. \ref{fig2:GL_E}.  Our first finding in the LDOS curves is a strong peak at zero energy, even for semiconductor ribbons, the so-called mid-gap state. This impurity-induced state is attributed to the repulsive short-range potential around monovalent impurities (atoms or molecules), which are strongly bonded to one of the carbon atoms in graphene.\cite{wehling2007local,wehling2009impurities,gonzalez2012impurity, Lehtinen2003,Ayuela2010,Santos2012} 
Generally, this mid-gap state could be related to the magnetic properties of the material, which it has been widely studied and is beyond the scope of this work.\cite{son2006energy}

Besides of the impurity-induced peak at $E=0$, we observe common features in the LDOS curves for all configurations in Fig. \ref{fig2:GL_E}. At energies below $0.4$ $\gamma_0$, defined by the first van Hove singularity of the pristine $N=11$ AGNR, the LDOS curves present a series of well-defined sharp peaks homogeneously spaced. 
The energy separation and the total number of those peaks are determined by an effective system length $L_{eff}(>L)$, defined as the mean distance between the doped regions, as it is shown in shaded on Fig. \ref{fig1:esq}. In this sense, the separation between I-configuration resonances is shorten than those of the D-configurations, since  $L_{eff}^{I} > L_{eff}^{D1 (D2)}$. These states get sharper and closer in energy as $L_{eff}$ is increased, and it is possible to follow continuously their evolution as the  system length is enlarged. If the ribbon width $N$ is increased, we can still find resonant levels in the low energy range similar to the presented in Fig. \ref{fig2:GL_E}, independently if the ribbons are metallic or semiconductor. For wider metallic ribbons, the low-energy range is minor; nevertheless, is still possible to observe the same patterns of LDOS. For semiconductor ribbons we have observed the impurity mid-gap state at $E=0$, and resonant levels above the in
 herent g
 ap of the ribbon. 

This quantum well-like behavior is, in fact, a hallmark of the resonances arising from confinement between the hydrogenated regions.\cite{xiong2013molecular,jw_PRB83} 
The regular periodicity of these states allows us to follow resonances above energies of $0.4$  $\gamma_0$, which are mixed with other kinds of energy levels. In the following sections, we will explain this LDOS behavior by using simple band-folding models.

In Fig. \ref{fig2:GL_E} it is also presented the two-terminal conductance of the system as a function of energy (red online). The hydrogen ad-atoms impose an extra spatial electronic confinement, which produces some resonant states between the doped regions, that act as open channels for electron transmission. The conductance profiles present a series of well defined peaks becoming from strongly quasi-localized states, which acquire a finite lifetime due to their indirect coupling through the continuum spectra of the leads, in analogy of the mechanism of formation of Bound States in the Continuum (BICs) in high symmetrical systems.\cite{JW_EPL10}

For the I-configuration, the doped regions present selective doping  where hydrogen ad-atoms occupy one of the grapehene sublattices. Such arrangement may arise, for instance, when long-chain hydrocarbon molecules are attached periodically on certain graphene sites.\cite{PhysRevB.89.045433,Santos2012,Gargiulo2014} This selective doping favors the hybridization between the resonant states and the continuum spectra of the leads, allowing a better conduction through the system in the low-energy range and even at higher energies. 
This clear resonant behavior, where the conductance curve presents well defined conduction energy channels, could be used for novel electronic applications, such as spin filter or molecular sensor devices.

In the figure \ref{fig2:GL_E} we observe that the conductance curves for the D1 and D2 configurations are clearly different in comparison with those of I-configuration. For low energy, although the sharp states are also present in the LDOS curves, the D1 and D2 configurations do not conduce because the 
C-H bonds in each doped region are balanced in both graphene sublattices (see Fig. \ref{fig1:esq}). 
The D1 configuration is a poor conductor (similar to an insulator) for all energies. This ad-atom distribution has pairs of contiguous carbon atoms with adsorbed hydrogen ad-atoms in different sublattices and, as a consequence, the electronic transport is mostly blocked. In fact, at low energies the effect on the conductance of the C-H bond is similar to an ideal vacancy since the $\pi$-type bonds are no longer available for conduction. \cite{Pereira_PRL96}
The  D2-configuration favors the electron transmission at energies over $0.4$ $\gamma_0$, which is defined by the bottom of the first parabolic conduction sub-band, as in the case of a line-defects in graphene.\cite{gunlycke2012confinement} In particular, for this configuration, the attached hydrogen atoms are separated by carbon atoms, therefore, localized states appear between the impurity positions which enhance the conductance through their hybridization with the resonant states of the system.

It is noteworthy that, for all configurations, the conductance curves above the first van Hove singularity are dominated by parabolic bands, well-known in semiconductors, so they can be outlined using a toy model for non-relativistic electrons incident on a double-barrier potential within the Schr\"odinger theory.\cite{harrison2005quantum}

\begin{figure}[t!]
\includegraphics[width=0.5\textwidth,angle=0,clip]{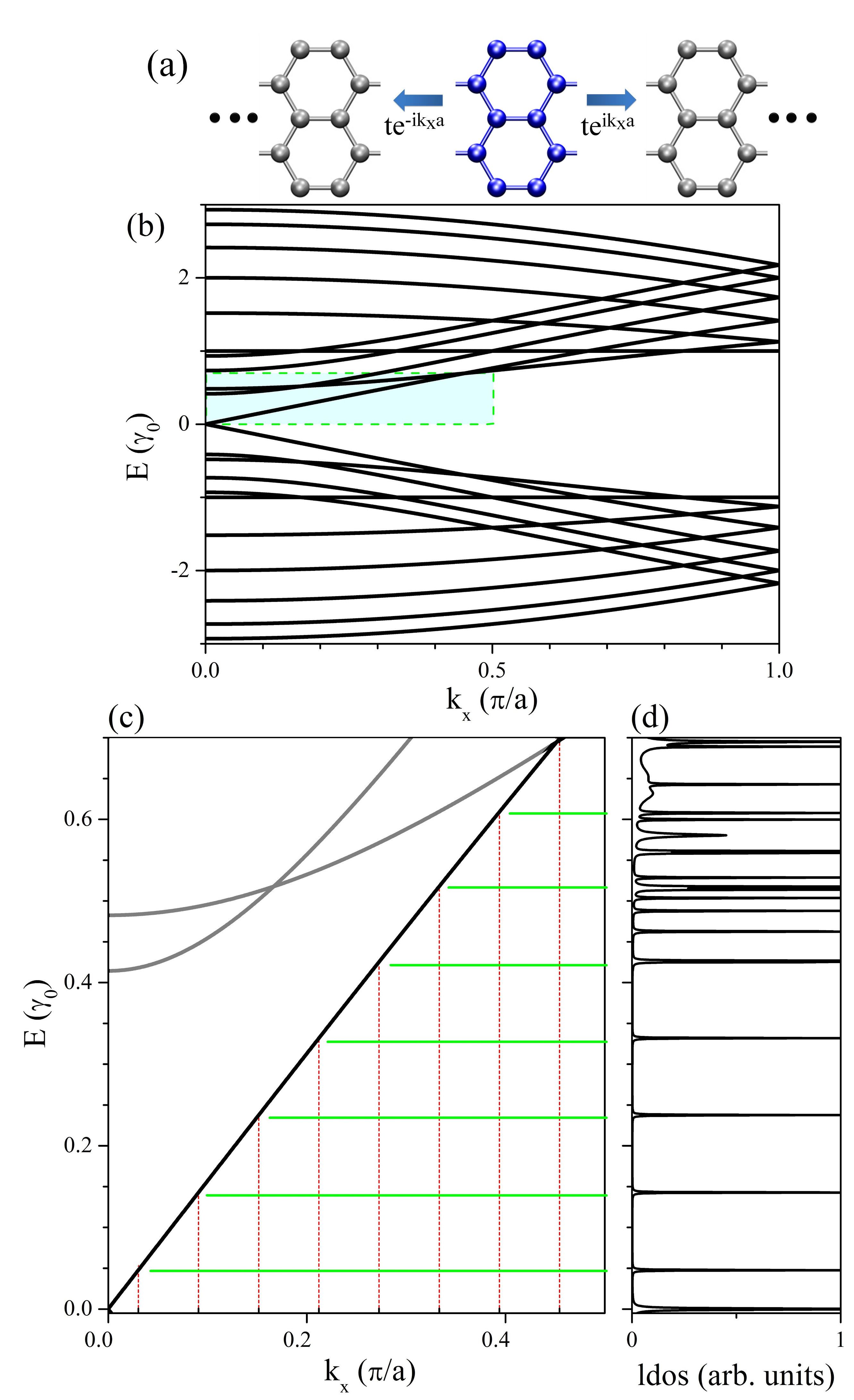} 
\caption{(Color online) 
(a) Model used for folding the electronic structure in the $k_x$-direction. 
The atoms in the unit cell are highlighted.
(b) Band structure of the pristine metallic ribbon $N = 11$. 
(c) Zoom of the low-energy bands with the $k_x$-space quantization for a length $L = 15$, 
(vertical dashed lines). The horizontal lines correspond to the energies where the first band  intersect  the quantization lines.
(d) LDOS of the D2 configuration for $L = 15$.}
\label{bands_kx}
\end{figure}

\subsection{Characterization of Resonances and Interface-states by Band-folding Models}
In this section, we unravel the tendencies of the LDOS curves shown in Fig. \ref{fig2:GL_E}. 
We use the electronic structure of simplified models, which are folded in the reciprocal 
space according to the width and length of the system.
We find a simple procedure to predict the energies of the resonances and their spatial 
distribution across and along the system. In what follows, we show results for the D2 distribution; nevertheless, other configurations can be equally explained with this approach. 

We start the analysis by explaining the sharp states exhibits in the LDOS curve of the considered system (Fig. \ref{bands_kx} (d)), which arise from the metallic band of the ribbon. We have used a simplified model of a pristine $N=11$ AGNR with the unit cell shown in Fig. \ref{bands_kx} (a), and the corresponding  band structure displayed in Fig. \ref{bands_kx} (b) and (c). 
Hydrogen ad-atoms impose different boundary conditions at both ends of the system. Regarding the doping distribution per sublattice at the ends, the D2-case present a crystallographic spatial inversion symmetry which determines the properties of the confined electron wave-function. The effective length $L_{eff}$ of the system is defined by the distance between the centers of the doped regions, which define the band folding projected along the ribbon axis.  For each sublattice, the confined states are similar to standing waves in a pipe with open end (details in Appendix \ref{Apex_Kx}), and they have a discrete set of $n$ allowed $k_x$-values, given by:
\begin{equation}
\vert k_x^n \vert= \frac{\pi}{L_{eff}} \left( n-\frac{1}{2} \right),\label{Eq:kxn}
\end{equation}
where $L_{eff}$ is the effective length (note that $L_{eff} \geq L $, presented as the shaded regions of Fig. \ref{fig1:esq}), and the $n$ index is a positive integer greater than zero. 

By using the Born-von Karman boundary conditions, the intersection of the allowed  $k_x^n$-values (vertical red online) with the envelope band (diagonal black online) determine the energies of the series of equidistant peaks in the LDOS curves of Fig. \ref{bands_kx} (d). Here, the horizontal lines (green online) indicate the energies of those intersections, which are completely aligned with the LDOS peaks. The spatial distributions of the lower metallic states are shown in the Appendix \ref{Apex_Kx}, which are in a good agreement with those  calculated by density functional theory (DFT). In this sense, we are testing our calculations against more sophisticated methods included in Appendix \ref{Apex_dft}. 

\begin{figure}[ht!]
\includegraphics[width=0.5\textwidth,angle=0,clip]{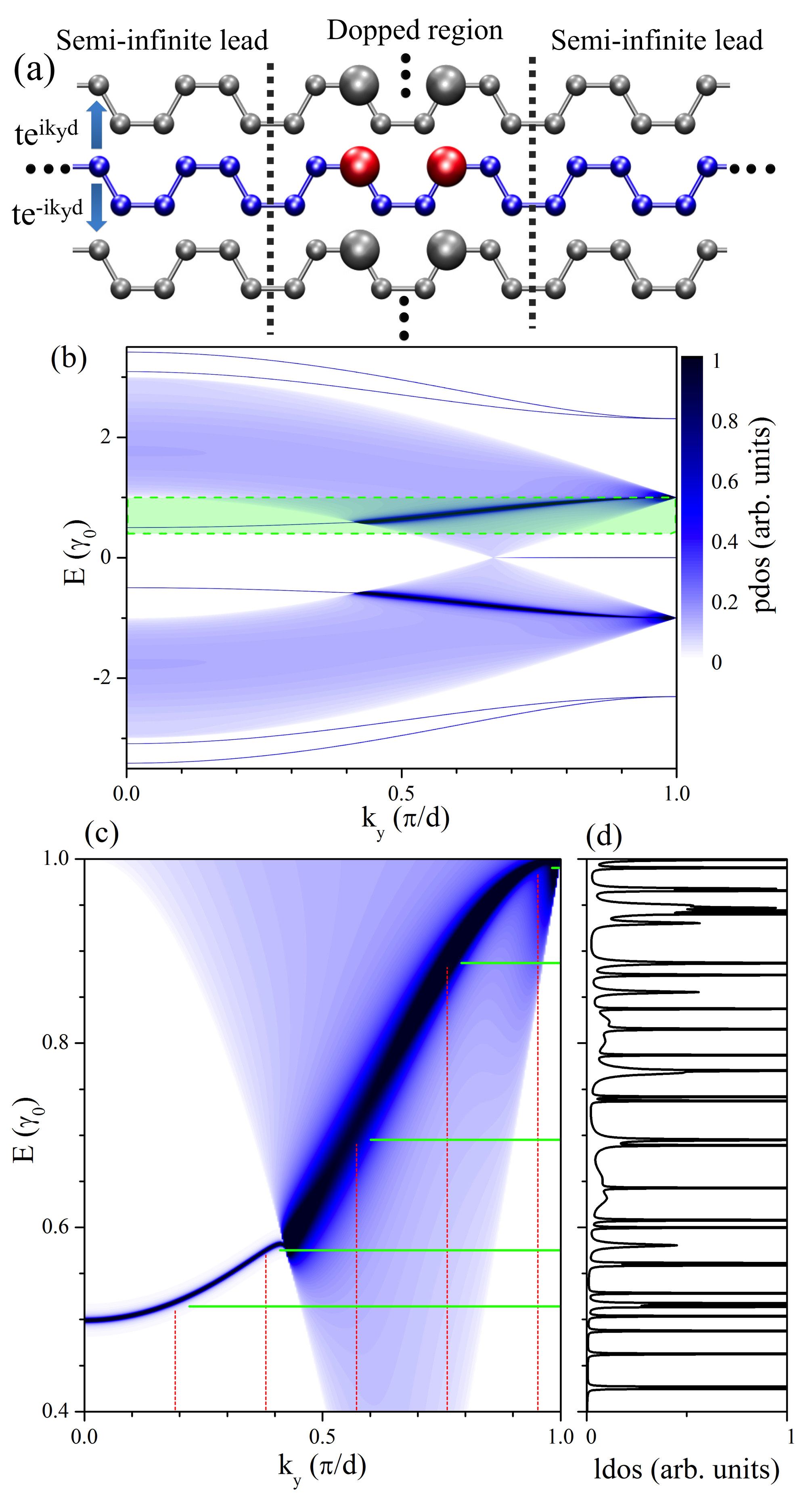} 
\caption{(Color online) (a) Effective model used for the band folding in the $k_y$-direction. 
(b) Projected LDOS (pdos) in the reciprocal space ($k_y$-direction). 
(c) Zoom of the pdos with the $k_y$-quantization lines (vertical lines) for a $N = 11$ AGNR.
The horizontal lines correspond to the energies where the projected pdos and the quantization lines intersect. 
 (d) LDOS of the D2-configuration for a $N=11$ AGNR and doped regions separated by a length $L = 15$.}
\label{bands_ky}
\end{figure} 

It is noteworthy to mention that we can apply the same folding procedure to semiconductor ribbons. We find that the $k_x$-space discretization also accurately predicts the energy at which the resonant sharp states appear. The difference with the metallic ribbons is that now these states are no longer equidistant, due to the curvature of the parabolic envelope band, which depends on the ribbon width. 
Even though resonant states are predicted with high accuracy, regardless of the character of the bands for any system length, the precision decreases with the number of sub-bands.

Although above the first van Hove singularity we can still identify states from the metallic envelope-band, the LDOS curves display other broader states which arise from the folding of parabolic sub-bands of higher energies or from the interface states of the system (within the doped regions). The interface states are identified by using an effective scheme, where each doped region of the system is individually modeled as an infinite line of defects forming a crystallographic discontinuity between two semi-infinite graphene sheets, as it is depicted in Fig. \ref{bands_ky} (a). 
Considering the unit cell of the effective model (highlighted in this figure with blue online) we have applied the Bloch's theorem in the perpendicular direction of the system with a lattice constant $d = \sqrt{3}\,\,a_{CC}$. We consider nearest-neighbors interactions in the vertical direction, with a $k_y$-dependent hopping parameter. Thus, we have calculated the $k_y$-projected density of states (pdos), given by:
$\mathcal{D} \left( E, k_y \right) = -\frac{1}{\pi} \mathrm{Tr}\, 
\left[\mathrm{Im}\, G_{C} \left( E, k_y \right)\right]$, 
where $G_{C}$ is the Green function defined in the Appendix \ref{Apex_cond}. 
Note that the $\mathcal{D} \left( E, k_y \right)$ now is $k_y$-dependent, as shown in Fig. \ref{bands_ky} (b) for the D2-configuration.

We observe the expected interface states at $E=0$ in the range $2/3 \leq |k_y d/\pi| \leq 1$. These flat bands, correspond to localized edge states and are associated with the characteristic zigzag ends of semi-infinite graphene sheet in the simplest non-interacting approach.\cite{fujita1996peculiar,wakabayashi2010electronic}  
Although these states around $E=0$ are not the main issue of this work, note that in DFT calculations they show spin polarization,\cite{son2006half} results which are the source of recent controversy due to the possible effects of spin contamination.\cite{arb1,arb2,arb3}

We also find an electron-hole symmetry of the  bonding and anti-bonding interface bands produced by the impurities, which lie between the graphene continuum of states.
At the $\Gamma$-point the splitting of these levels is approximately $\gamma_0$, and it decreases as the horizontal distance between the C-H pairs within the barrier is increased. 
This energy splitting of the interface states only appears when the sublattices are locally balanced with the same number hydrogen pinned at each graphene sublattice, otherwise, the interface bands are not split, and the states remain at $E = 0$, extending from $|k_y d/\pi|=0$ to $1$.

For AGNRs, applying the folding procedure in the perpendicular direction generates a set of $n$ allowed $k_y^n$-values (vertical lines in Fig. \ref{bands_ky} (c)), which are given by: 
\begin{equation}
\vert k_y^n\vert = n \frac{\pi}{W} ,
\end{equation}
where $n =1,\,2,\,...$ and $W = \frac{\sqrt{3}}{2}$ $\left( N-1\right)$ $a_{CC}$ corresponds 
to the ribbon width. The intersection between the obtained $k_y^n$-values with interface bands determines the energy at which these states appear in the LDOS curves of the system, as it is shown in Fig. \ref{bands_ky} (c) and (d).
As the interface bands coexist with the graphene continuum states, for $k_y \geq 0.4 $, they become broader indicating shorter lifetimes. Due to the hybridization between these states (Fig. \ref{bands_ky} (c)), the electron conduction at these energy levels is allowed, bottom panel of Fig. \ref{fig2:GL_E}. The spatial distribution of interface states is presented in the Appendix \ref{Apex_Kx}. 
These interface states appear in other carbon-based nanostructures such as grain boundaries in graphene and jucntions in carbon nanotubes.\cite{Chico_PRB80,pelc2013grain,ayuela2008electronic} It is worth mentioning that these interface states, usually considered as a weakness in device performance, may provide a method for achieving diode behavior at the nanoscale.\cite{rochefort2002quantum}

\subsection{Dependence on the Ribbon Width}\label{Any_Width} 
\begin{figure}[h!]
\includegraphics[width=0.49\textwidth,angle=0,clip]{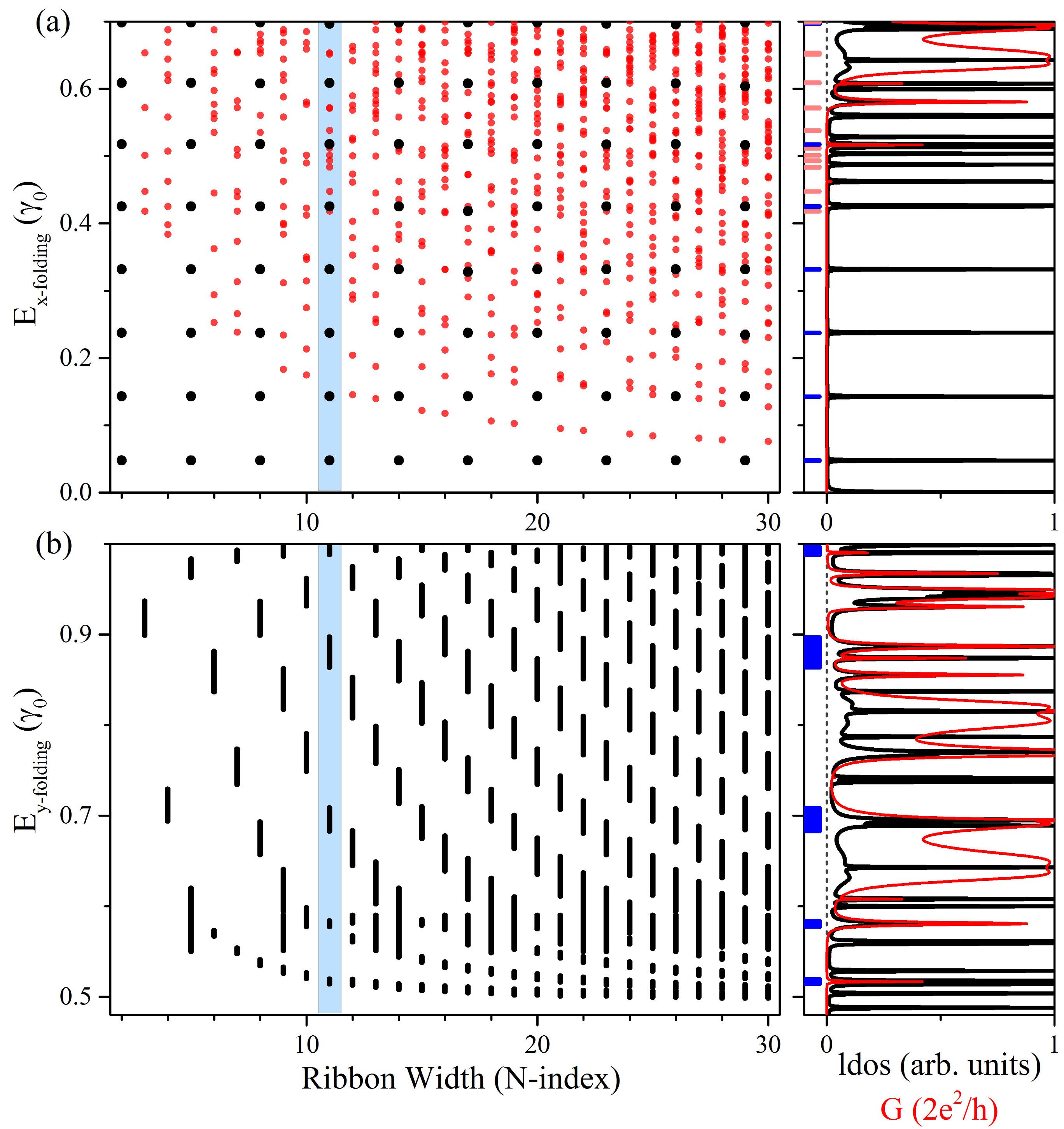} 
\caption{(Color online)  (a) Energy position of the intersection of the bands and the $k_x$-values as a function of the AGNR width, for a system length $L = 15$
(b) Interface states arising from the $k_y$-folding for any AGNR width. 
As a comparison, we have included the LDOS and conductance for the width $N=11$ and a system length $L = 15$ (shaded region).}
\label{fig:anywidth}
\end{figure}

We focus on the effects of the AGNR width on the resonant and interface-states. We choose a system length $ L  =  15$ for which we have applied the folding procedure in the $k_x$-direction in Fig. \ref{fig:anywidth} (a). For metallic AGNRs, energy states stemming from the linear metallic band are shown in black marks in this panel. These states are regularly distributed in energy determined by the system length $L$.
For the parabolic sub-bands, the allowed states are given by red marks, and we identify them with some peaks of LDOS. In the upper-right panel, we indicate with blue bars the energy of the states arising from the metallic envelope-band and with red bars those states emerging from the folding of the parabolic sub-bands.
Note that for an AGNRs width of $N=11$, shadowed in left panels, there is a good agreement between the LDOS and the metallic resonances obtained by the folding procedure. Besides, states coming from parabolic bands are more difficult to follow since they are spread in the doped regions and, as a consequence, their energies are shifted. 

By applying the folding procedure in the $k_y$-direction, we identify the energies of the interface states as a function of the AGNR width, as shown in Fig. \ref{fig:anywidth} (b). 
These interface states are observed for energies greater than  $0.5 \, \gamma_0$ independent of the system width. Above $0.6 \, \gamma_0$ in the energy range of the hybridization between interface states and continuum bands of graphene, the interface peaks become broader and contribute to the conductance.
In the bottom-right panel, we compare the LDOS and conductance with the blue marks belonging to the $k_y$ band-folding for an AGNR of width $N=11$.

\subsection{Comparison with Experimental Conditions} 
In this section we aim to get into contact with the experiments; therefore, we focus on how these quasi-localized, resonant and interface states behave in gapped graphene or under random distributions of adsorbed hydrogen at low concentration.

\begin{figure}[ht!]
\includegraphics[width=0.47\textwidth,angle=0,clip]{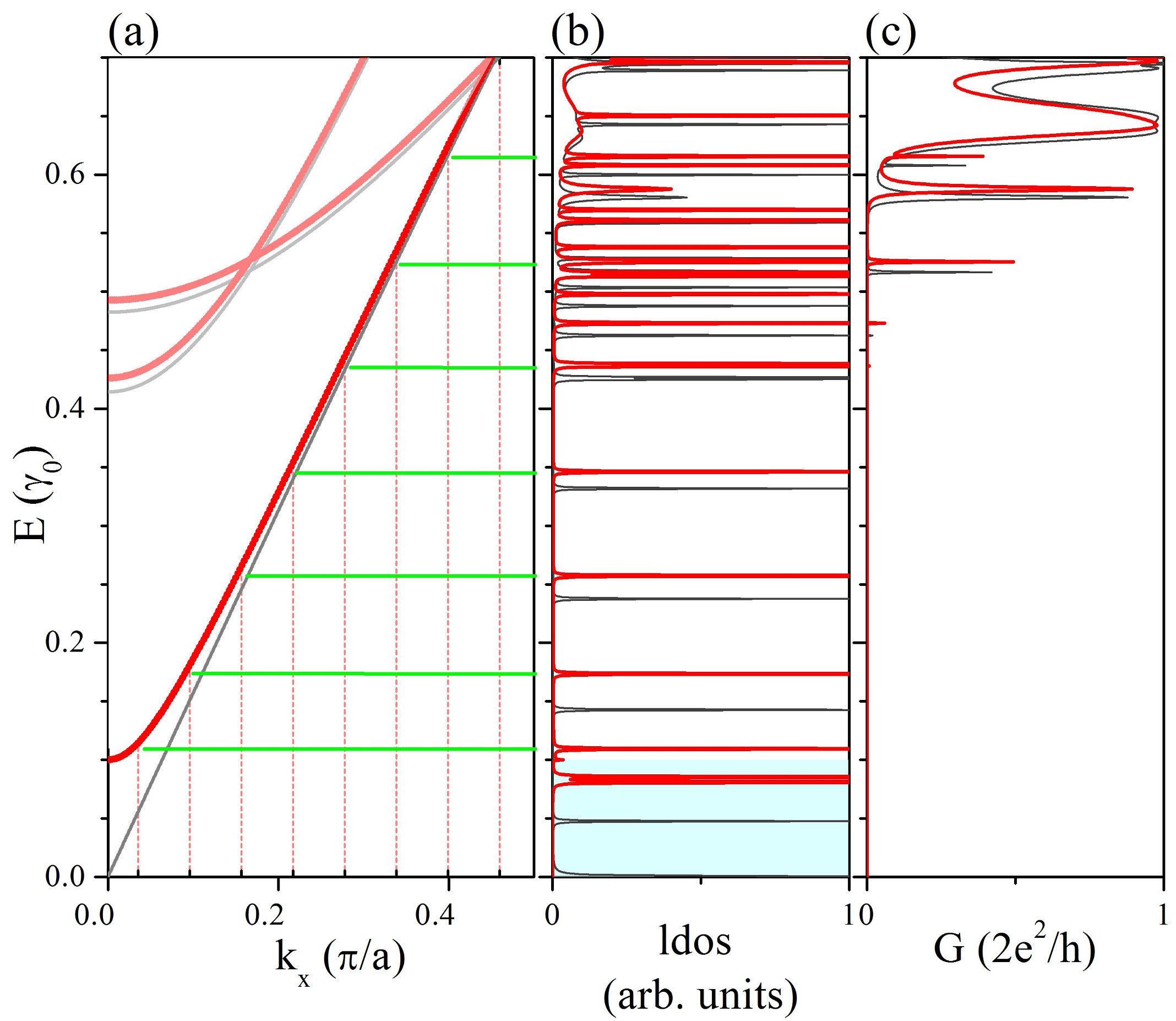} 
\caption{(Color online) (a) Low energy bands for gapped graphene (red online), the $k_x^n$ quantization for the effective length L = 15, has been included as vertical dashed lines. 
The horizontal lines determine the energies where the first band of the gapped case and 
the quantization lines intersect, 
(b) the density of states and 
(c) two terminal electronic conductance of the D2 configuration for a separation $L = 15$.
For comparison, the gapless case has been included by black lines in all the panels.}
\label{fig:mass}
\end{figure}

\subsubsection{Gapped Graphene}
In previous sections, we have focused on the electronic properties of systems based on 
nanoribbons made of gapless graphene. However, most applications in electronics require a band gap, so a some actual devices are manufactured over gapped graphene. 
The interaction of the graphene layer with the substrates, such as hexagonal boron nitride (hBN) or silicon carbide (SiC), opens band gaps, which depend on the lattice mismatch and the relative angle between the graphene and  substrate lattices.\cite{zhou2007substrate,giovannetti2007substrate,jung2014origin,ortix2012graphene}
Additionally, two-dimensional materials with hexagonal symmetry such as hBN, silicene or germanium can be described by using a graphene-like tight-binding Hamiltonian with an additional mass term.\cite{ribeiro2011stability,liu2011low,guzman2007electronic}
In this context, we model the gapped graphene by considering a constant staggered potential, namely, a local potential that acts with the opposite sign in the two sublattices.  In this model, the mass of the electrons is a parameter corresponding to the difference $\vert V_{A}-V_{B} \vert$, \cite{gonzalez2012impurity}  which in the case of graphene over a SiC substrate correspond to $V_{A(B)} = \pm 0.1 \gamma_0$.\cite{zhou2007substrate}  

The electronic states of the system are modeled by using the same unit cell described in Fig. \ref{bands_kx} (a). By applying the folding procedure along the $k_x$-direction it is possible to calculate a set of allowed discrete values $k_x^n$ defined in Eq. (\ref{Eq:kxn}). Note that these discrete momentum values are the same of the gapless graphene system, because the mass term is diagonal, and does not affect the translational symmetry of the system.
In Fig. \ref{fig:mass} (a), for  metallic $N = 11$ AGNR, we compare the energy bands of gapless and gapped graphene, which are denoted by the black and red lines respectively. 
The main effect of the mass term is to transform the previously linear metallic band into the first parabolic band, which defined the gap of the system.
Near the Fermi level, the energy distance between the LDOS peaks, determined by the folding procedure, changes because the envelope band is no longer linear, as seen in Fig. \ref{fig:mass} (b).
At higher energies, the separation between the peaks of the LDOS gets close to previous values. The conductance curves are slightly affected since the effects of the mass term are mostly observable near the Fermi level, Fig. \ref{fig:mass} (c).
The impurity induced mid-gap state previously located at $E = 0$ is now split in to series of peaks within the gap defined by the staggered potential. These states preserve the electron-hole symmetry only when the doped sublattice balance is conserved. We find that these defect states present sublattice localization in the doped region, in the vicinity around the hydrogen ad-atoms. 
All the main characteristics of these in-gap peaks, such as energies, degeneracy and spatial 
distribution for gapped graphene, can be inferred in detail by using band-folding analysis similar to the presented before in Fig. \ref{bands_ky}. 

\begin{figure}[hb!]
\includegraphics[width=0.47\textwidth,angle=0,clip]{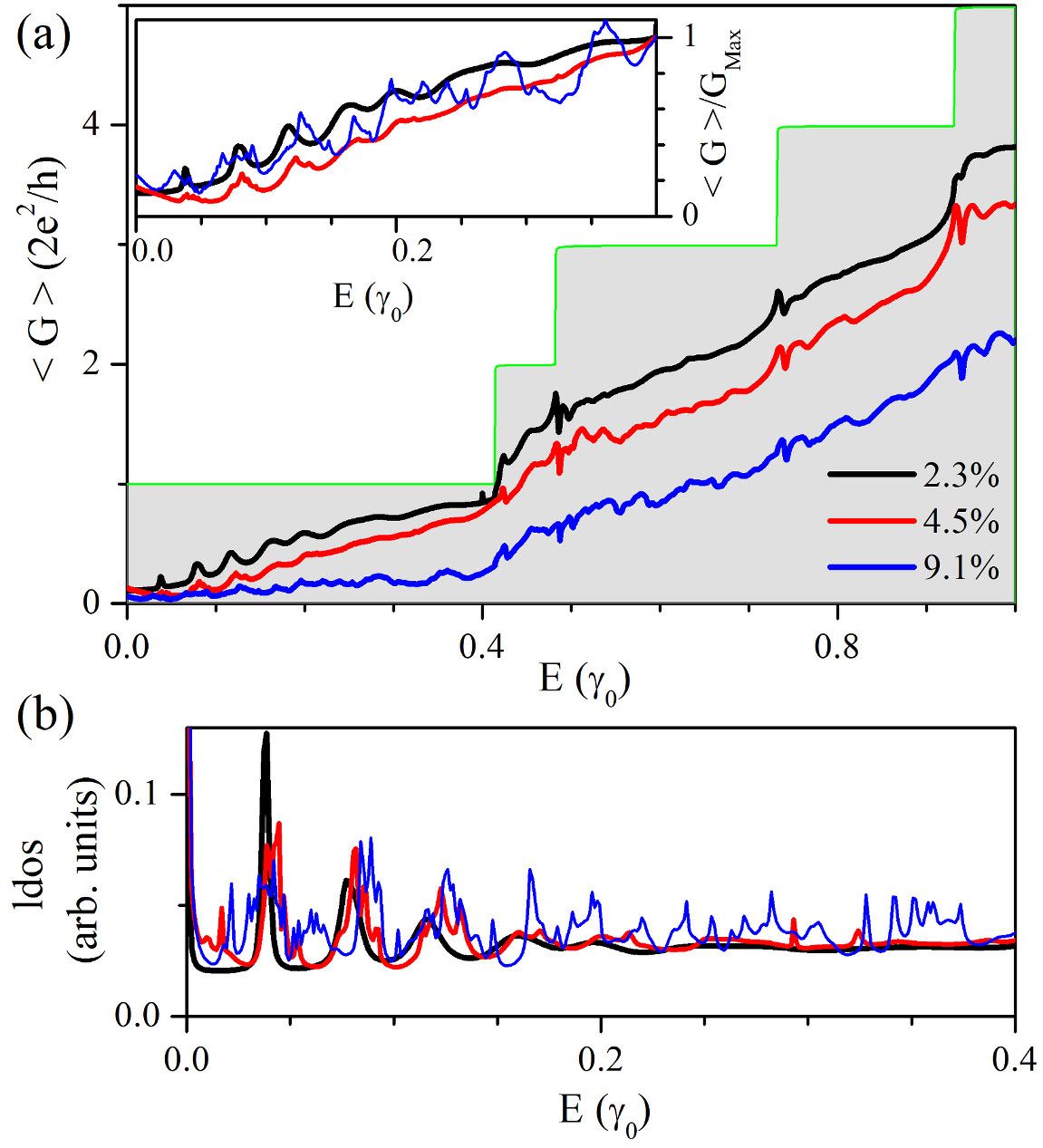} 
\caption{(Color online) (a) Average conductance and (b) average LDOS as a function of the energy for a $N=11$ AGNR separated by $L=15$ for several hydrogen coverages.
Black lines indicate H coverage per doped region of 2.3 \%; red, 4.5 \%; blue, 9.1 \%.
The conductance of the pristine case is shadowed. The inset of the (a) panel includes the normalized  conductance at the first van Hove singularity.}
\label{figG}
\end{figure}

\subsubsection{Randomly Adsorbed Hydrogen at Low Concentrations}
We wonder about how robust is the effect of confinement in the presence of a random distribution of H impurities. We consider low concentrations of H ad-atoms randomly distributed in the two doped regions. The percentage of doping is defined as the ratio $(N_H/N_C)\times100$, where $N_H$ is the number of H ad-atoms and $N_C$ is the total number of C atoms in the doped region. Finally, we  impose that the number of hydrogen ad-atoms is the same in each doped region.

Figure \ref{figG} shows the conductance and LDOS for the $N = 11$ and  $L = 15$ AGNR as a function of the Fermi energy, averaged over 100 independent random realizations for low coverage. The concentration of random hydrogen doping strongly affects the electronic properties. As the concentration increases, the conductance of the systems rapidly decreases due to the disorder generated by the random adsorption at the doped regions.

We note that at low energy and low concentration of H-atoms, there are well-defined energy levels in averaged LDOS and conductance curves, shown in Fig. \ref{figG} (a) and (b). These states resemble to those belonging to the ordered configurations; however, the energy separation between them is the half separation of the previous results, which is related with the statistical correlation of the considered disorder. We find that the number of resonance peaks increases with the average separation between doped regions. Since the low-energy oscillations appeared both in conductance and LDOS, it seems that scanning tunneling spectroscopy could be used to detect those oscillations, as they were already measured in carbon nanotubes.\cite{Buchs2009}

Additionally, regardless of the distance between the doped regions, we observe the Fano-like resonances\cite{JW_EPL10,gong2013line} in the average conductance, at energies corresponding to the van Hove singularities of the pristine system. This effect could be used to determine indirectly the width of the ribbons employed in the system.

\section{Final Remarks}
In summary, we have shown that is possible to confine electrons in a GNR-based system. 
We accomplish  that, by theoretically studying hydrogen-like ad-atoms in two separate regions. Our results show different types of electronics states, such as resonant and interface states, which determine the transport properties of the system. We have found evidence of the induced quantum well-like behavior in case of gapped graphene or regardless of the hydrogen distribution in the doped regions, whether they are highly ordered or diluted random distribution.  

We are able to predict the energies and the spatial distribution of these electronic states by applying a simple band-folding scheme to some basic models. For the resonant states, we have apply the Born-von Karmarn boundary conditions to the metallic envelope-bands of the pristine AGNR. Thus, the energies of the resonant states are equally spaced.  Furthermore, this band-folding scheme applies to the parabolic envelope band in both gapped graphene and semiconductor ribbons.
Perpendicular to the doped ribbon, the folding scheme drives to interface states separated in two groups: states that hybridize or not with the continuum band. We found that hybridized states contribute with broad peaks to the electronic conductance.
These proposed band-folding was successfully compared with density functional theory calculations.

Our results would encourage experimentalist to make these kind of systems and characterize their quasi-localized states, for instance, employing scanning tunneling spectroscopy. 
The obtained resonant electronic transport behavior could be externally manipulated by applying gates and/or side potentials, which allows to control the conductance of the system. This fact can be used in novel electronic applications and molecular sensor devices.

This work has been partially supported by the Project FIS2013-48286-C2-1-P of the Spanish Ministry of Economy and Competitiveness MINECO (JWG, AA).
Chilean FONDECYT grants, 1140388 (LR), and 1151316 (MP), and DGIP/USM internal grant 11.14.68 (MP), CONICYT ACT 1204 (LR, MP). JWG gratefully acknowledges the hospitality of the UTFSM (Chile) and fruitful discussion with L. Chico ICMM-CSIC (Spain).



\appendix

\section{Calculation of Conductance}\label{Apex_cond}
We include the technical details concerning to the calculation of the LDOS and the two terminal conductance using the Landauer formula and the surface Green's function matching formalism.\cite{Nardelli_1999,Datta_book,jw_PRB} 
This method divides the system in three blocks. 
There are two semi-infinite leads made of AGNR, left L and right R described with $H_L$ and $H_R$ Hamiltonian respectively.   
The finite central part is described by the Hamiltonian matrix $H_C$, which includes the two doped regions and the separation between them, as depicted in Fig. \ref{fig1:esq}. 
Thus, the total Hamiltonian is given by:
\begin{equation}
H= H_C  + H_R + H_L + V_{LC} + V_{RC},
\end{equation}
where $V_{LC}$, $V_{RC}$ are the coupling matrix of the left $L$ and 
right $R$ leads with the central region. 
The main objective in the method is to calculate the Green's function, which can be written as:
\begin{equation}
G_C(E) = (E \hat{I} - H_C - \Sigma_L -\Sigma_R)^{-1},
\end{equation}
where $\hat{I}$ is the identity matrix, $\Sigma_\ell= V_{\ell C}\; g_\ell \;  V_{\ell C}^\dagger$ is the
self-energy of each lead $\ell=L,R$, and $g_\ell = (E -H_\ell)^{-1}$
is the renormalized Green's function of the semi-infinite lead $\ell=L,R$. 
In the linear response regime, the conductance $G$ is calculated as a function of the Fermi energy $E$, within the Landauer formalism. In terms of the Green's function for the 
system,\cite{Datta_book,Nardelli_1999} $G$ reads:
\begin{equation}\label{LandauerG}
G(E) = \frac{{2e^2 }}{h}T\left( {E } \right) = \frac{{2e^2 }}{h}
{\mathop{\rm Tr}\nolimits} \left[ {\Gamma _L G_C \Gamma _R
G_C^\dagger} \right],
\end{equation}
where  $T\left( {E } \right)$ is the transmission function across
the conductor, and  $\Gamma_{\ell}=i[ {\Sigma _{\ell}  - \Sigma
_{\ell} ^{\dag} }]$ is the coupling between the conductor and the leads.

Finally, within the Green's functions formalism, the LDOS per atom is proportional to a Green's function matrix element,\cite{economou1984green} in a particular the LDOS at the atomic position $j$, can be expressed as $\mathcal{D}_{j} \left( E \right) = -\frac{1}{\pi} \mathrm{Im}\, G^{j,j}_{C} \left( E \right)$, in such a way that the system LDOS is calculated by $\mathcal{D} \left( E \right) = -\frac{1}{\pi} \mathrm{Tr}\, 
\left[\mathrm{Im}\, G_{C} \left( E \right)\right]$ $=\sum_{\text{all atoms}} \mathcal{D}_{j} \left( E \right)$.

\begin{figure}[hb!]
\includegraphics[width=0.48\textwidth,angle=0,clip]{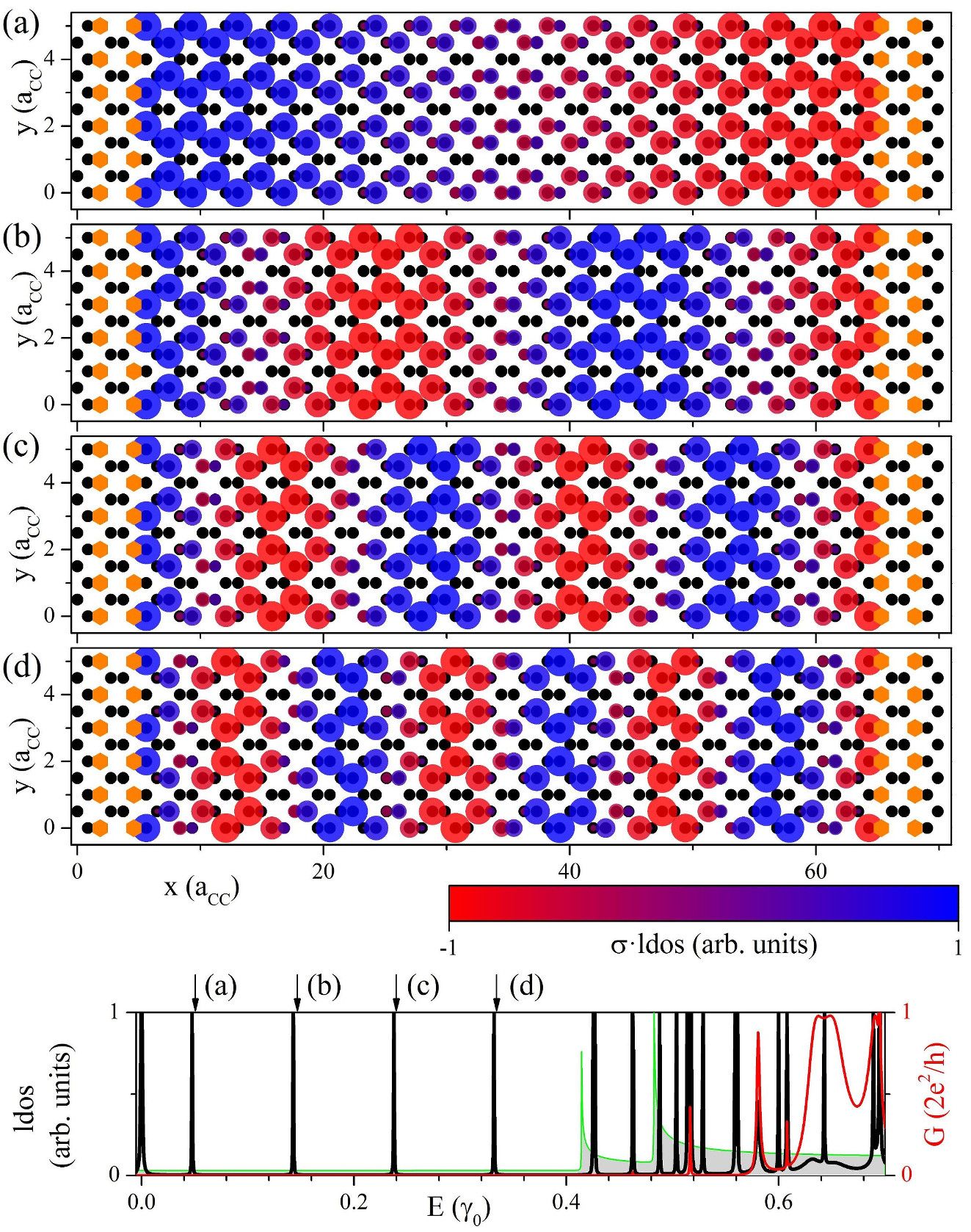} 
\caption{(Color online) LDOS per atom, red and blue colors represent the sublattice of the atom.
The bottom panel shows the LDOS (black line) and the conductance (red line) for a D2-configuration on a metallic AGNR $N = 11$ separated by $L = 15$. 
The arrows label the energies of the LDOS peaks plotted in the panels (a-d).}
\label{map_kx}
\end{figure}

\section{Spatial Distribution of Localized and Resonant States}\label{Apex_Kx}
\begin{figure}[t!]
\includegraphics[width=0.42\textwidth,angle=0,clip]{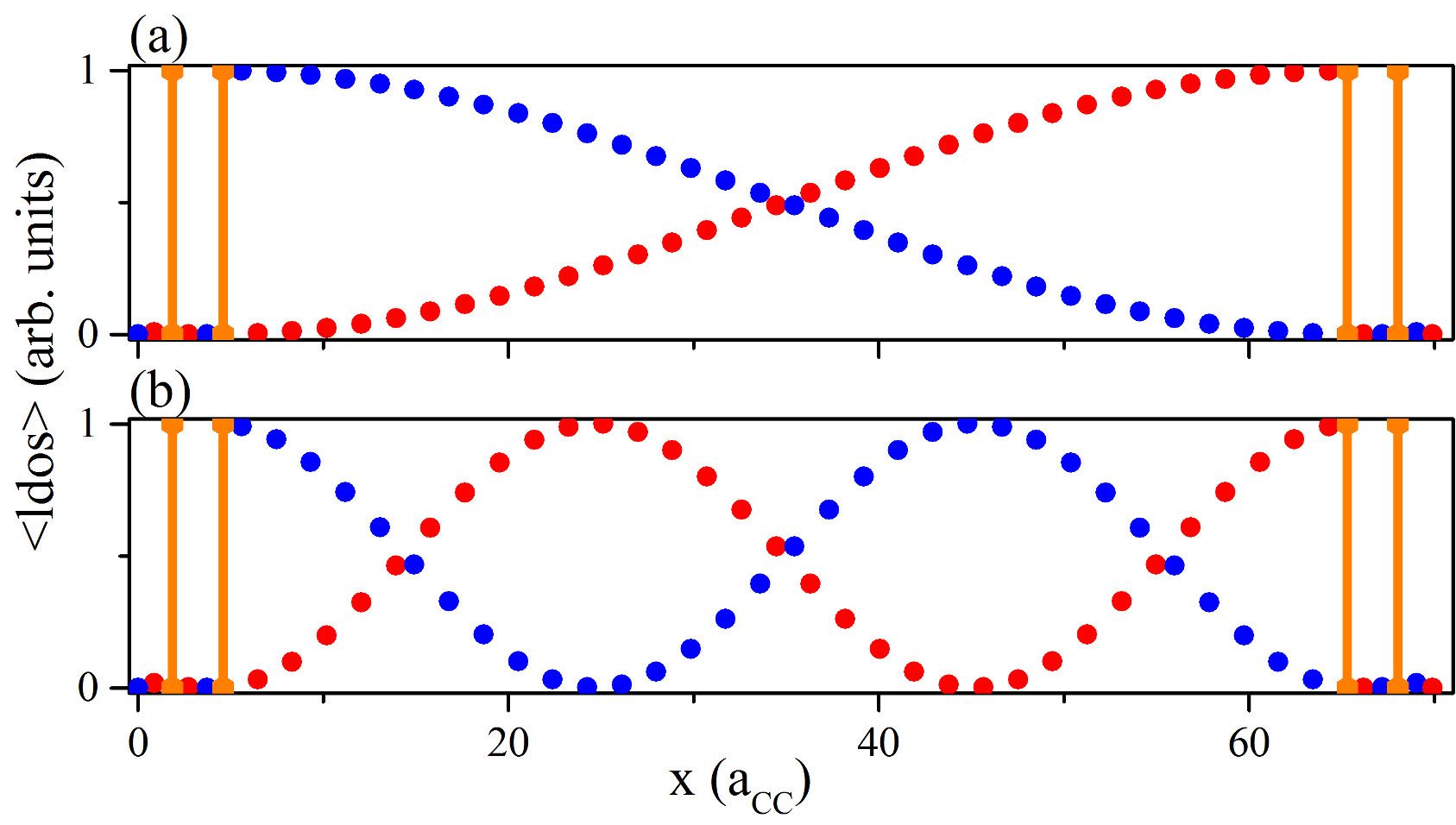} 
\caption{(Color online) Normalized average LDOS per column of atoms as a function of the $x$-position for the first two quantized $k_x^n$-values; (a) $n=1$ at $E = 0.048$ $\gamma_0$ and (b) $n=2$ at $E = 0.14$ $\gamma_0$.
Red and blue colors indicate the sublattice, and the orange lines indicate the position of  the hydrogen atoms on the ribbon.}
\label{fig:prof}
\end{figure}

We present the spatial distribution of the LDOS per atom where the red / blue color indicates different sublattice, for the lower sharp states belonging to the metallic band of the ribbon, labeled by (a-d) in the bottom panel of Fig. \ref{map_kx}. The bar color scale represents the density probability per sublattice.
It is possible to observe some general behavior for these states. First, there is an odd sequence of nodes along the longitudinal ribbon direction, defined by the allowed $k_x^n$-values. Second, the boundary conditions imposed by the impurities determine the spatial distribution of the LDOS for each sublattice, which exhibits the same feature of standing waves in a pipe with an open ends. This kind of behavior is similar to the observed in edge states of zigzag ribbons. To emphasizes these LDOS features, in Fig. \ref{fig:prof} shows the normalized average LDOS per column of atoms ($y$-direction) as a function of the $x$-position for the first two quantized $k_x^n$-values over the envelope metallic band (corresponding to panels (a) and (b) of Fig. \ref{map_kx}).

\begin{figure}[bh!]
\includegraphics[width=0.48\textwidth,angle=0,clip]{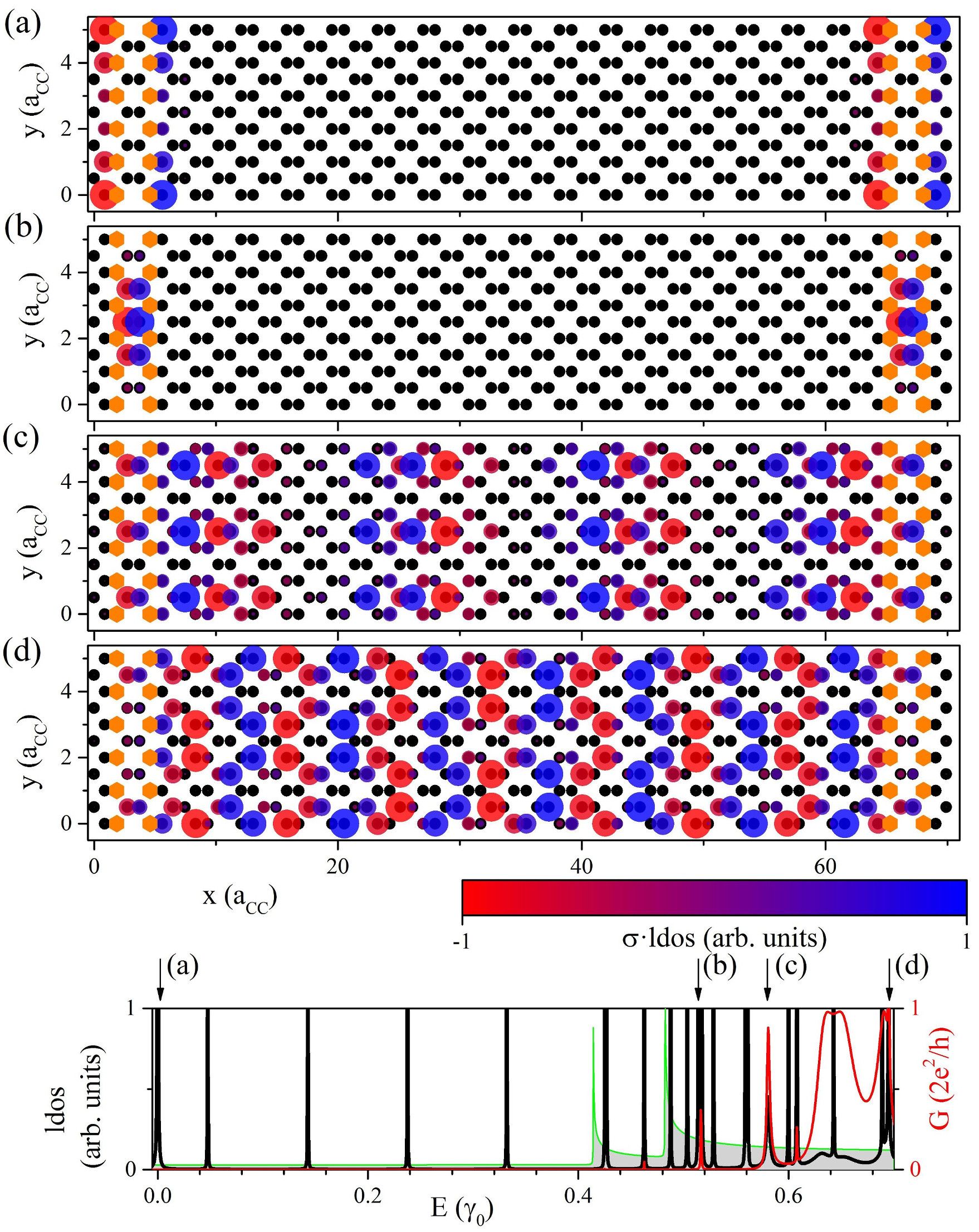} 
\caption{(Color online)  LDOS per atom of: (a) the zero-energy state and 
(b-d) the interface states at high energies, for the same ribbon parameters of Fig. \ref{map_kx}. Red and blue colors indicate the two graphene sublattices.}
\label{map_ky}
\end{figure}

Figure \ref{map_ky} shows the spatial distribution of the LDOS per atom corresponding to the energy states labeled by (a-d) at the bottom panel of the figure. The (a) panel shows the mid-gap state located at zero energy, present in all ordered configurations, regardless if the ribbon is metallic or semiconductor.\cite{wehling2007local,wehling2009impurities,gonzalez2012impurity} This impurity induced state is highly localized in the nearest-neighbors sites to the adsorbed hydrogen atoms. 

The (b) panel shows a non-conductive interface state, which is determined by the band-folding in the $k_y$-direction. This state corresponds to the first allowed discrete wavenumber, which is below the continuum of the graphene, and consequently, has a higher lifetime. This state is distributed between the hydrogen ad-atoms inside the doped regions.

In panels (c) and (d) show two resonant and extended states, determined by the band-folding in the $k_y$-direction, which are hybridized with the continuum of the graphene, shown in Fig. \ref{bands_ky}. The main feature of these states is that they are spread in a narrow energy window, so any electron from the leads injected at these energies easily passes across the system.\cite{JW_EPL10,deng2014formation}

\begin{figure*}[bh!]
\includegraphics[width=0.97\textwidth,angle=0,clip]{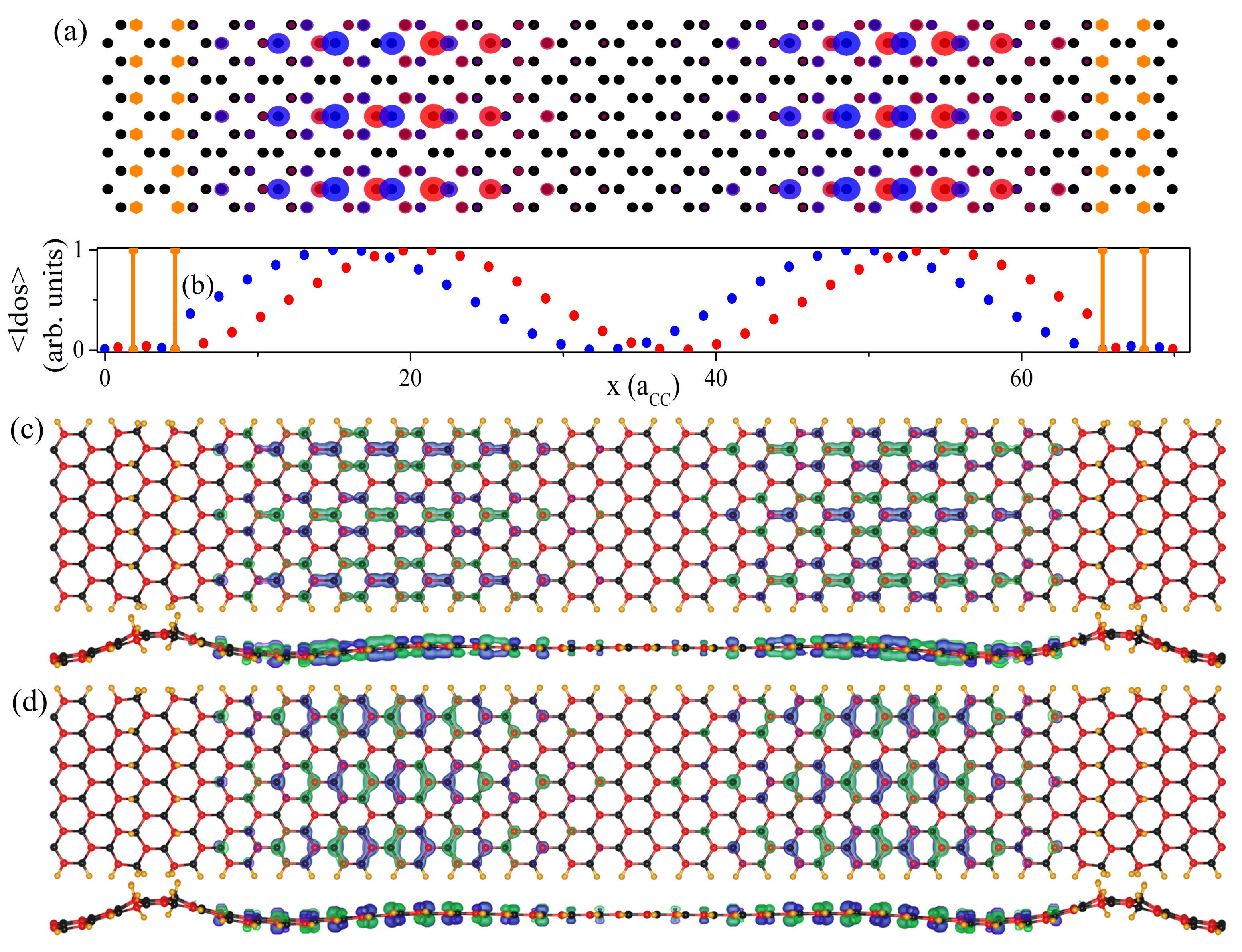} 
\caption{(Color online)  Comparison between the LDOS per atom (red and blue colors represent the sublattice of the atom) calculated within the tight-binding model and the wavefunction at $\Gamma$-point calculated within the DFTB+ for a parabolic-band state.
The (a) panel shows LDOS by the tight-binding model at $E = \pm 0.46$ $\gamma_0 \sim 1,2$ eV, and in panel (b) the averaged LDOS over the $y$-axis as function of the $x$-position. Red and blue colors indicate each graphene sublattice. The equivalent DFTB+ wave-functions correspond to  $E= 1.15$ eV and $E= -1.18$ eV in panel (c) and (d), respectively.}
\label{fig:TB_DFT_p}
\end{figure*}

\section{Comparison with Density-Functional-Theory Calculations}\label{Apex_dft}
The DFT calculations are performed within the superlattice approach using the density functional based tight-binding method implemented by the DFTB+ code\cite{aradi2007dftb+} with the associated Slater-Koster parameters.\cite{kohler2001theoretical} 
DFTB+ provides an efficient quantum simulation tool based in DFT, 
which allows us to calculate, in affordable times, the electronic properties of a system 
with the same size as calculated within a tight-binding approach. 
Self-consistent charge calculations are converged up to a tolerance of $10^{-8}$ e for the unit cell similar to the one shown in Fig \ref{fig:TB_DFT_p}. 
Ribbons are repeated periodically using super-cell approximation and are separated by an empty space of $10$ \AA $\,$ in the perpendicular directions. 
This cell is large enough to converge results with a single $\Gamma$-point. 
All atoms are relaxed within the conjugate gradient method until forces have been converged with a tolerance of $10^{-3}$  eV/\AA. We check that the relaxed geometry of a single hydrogen atom adsorbed on graphene reproduces well-known results.\cite{duplock2004hallmark,balog2010bandgap,boukhvalov2008hydrogen,chernozatonskiui2007superlattices} 
In agreement with these previous works, we found that the hydrogen atoms 
bonded to carbon atoms are about $1$ \AA\ over the graphene plane and its distance to nearest neighbor carbons is about $0.08$ \AA.

In Fig. \ref{fig:TB_DFT_p} we show the isosurfaces of the second harmonic-like state calculated by using the tight-binding and the DFTB+ approach. The latter includes the geometrical corrugation induced by the $sp^2$ to $sp^3$ bonds of carbon atoms attached to hydrogen. These states, calculated by two methods, have similar spatial distribution. 
In panel (a) we have plotted the spatial distribution of the LDOS per atom at $E = \pm 0.46$ $\gamma_0 \sim \pm 1.2$ eV obtained by the tight-binding method. 
As a comparison in Fig. \ref{fig:TB_DFT_p} (c) and (d), we have plotted the spatial distribution of the corresponding wavefunction (at energies $E= 1.15$ eV and $E= -1.18$ eV) obtained by using DFTB+ calculations. To facilitate the comparison, atoms have been colored in black and red according the corresponding sublattice. It is important to mention that the energies of the latter states do not preserve the electron-hole symmetry due to the electronic correlation effects; however, they still have a good agreement with the ones previously obtained in the tight-binding model.

Additionally, in the panels (c) and (d), a change in the sign of the wavefunction is evidenced around of the nodes. The wavefunction in (c) is even, while in (d) it is odd. The sign of resonant DFTB+ wavefunctions changes from bonding-like to antibonding-like character for negative and positive energies from $E_f$, respectively. 
This parity change means that optical elements between both resonances will be certainly far from zero, and optical transitions between pairs of states could be also detected in experiments. 

The wave-functions arising from parabolic bands remind standing-waves on a rope, where both sublattices exhibit nodes at both ends, as shown in Fig. \ref{fig:TB_DFT_p}(b).
This spatial distribution is different from the wave-function originated in the linear envelope-band of Fig. \ref{fig:prof}, where each sublattice presents a node at one end and a maximum in the other.

Finally, in DFTB+ calculations we look at the energy separation between resonant states. To the separation, we associate an effective energy scaling $\gamma_{\mathrm{eff}}$, taking in to account the resonant levels far from the Fermi energy. We obtain a value of $\gamma_{\mathrm{eff}} = 2.57$ eV, in good an agreement with the one used through this work.


\end{document}